\documentclass[journal,12pt,onecolumn,draftclsnofoot]{IEEEtran}
\usepackage{times}
\usepackage[final]{graphicx}
\usepackage{textcomp}
\usepackage[reqno]{amsmath}
\usepackage{amsfonts}
\usepackage{times,amsmath,epsfig}
\usepackage{latexsym,amssymb}
\usepackage{cite}
\usepackage{psfrag}
\usepackage{subfig}
\usepackage{stfloats}
\usepackage{amsmath}
\usepackage{amssymb}
\usepackage{array}
\usepackage[noend]{algorithmic}
\usepackage{algorithm}
\usepackage{bm}
\usepackage{color}
\usepackage{multicol}
\usepackage{multirow}
\usepackage{setspace}
\usepackage{tabularx}
\usepackage[T1]{fontenc}

\newtheorem{rem}{Remark}
\newtheorem{theo}{Theorem}
\newtheorem{prob}{Problem}
\newtheorem{assu}{Assumption}
\newcommand{\PreserveBackslash}[1]{\let \temp =\\#1 \let \\ = \temp}
\newcolumntype{C}[1]{>{\PreserveBackslash\centering}p{#1}}
\newcolumntype{R}[1]{>{\PreserveBackslash\raggedleft}p{#1}}
\newcolumntype{L}[1]{>{\PreserveBackslash\raggedright}p{#1}}

\newlength{\figwidth}
\begin{document}

\title{Energy-efficient Task Offloading for Cooperative Mobile Edge Computing under Sequential Task Dependency}
\author{
Xiang Li, Rongfei Fan, and Han Hu
\thanks
{
%
X. Li, R. Fan, and H. Hu are with the School of Information and Electronics, Beijing Institute of Technology, Beijing 100081, P. R. China. (\{lawrence,fanrongfei,hhu\}@bit.edu.cn,).
}
}

\maketitle

\begin{abstract}
In this paper, we study a mobile edge computing (MEC) system in which the mobile device is assisted by a base station (BS) and a cooperative node. The mobile device has sequential tasks to complete, whereas the cooperative node assists the mobile device on both task offloading and task computation. In specific, two cases are investigated, which are 1) the cooperative node has no tasks to complete itself, and 2) the cooperative node has tasks to complete itself. Our target is to minimize the total energy consumption of the mobile device and the cooperative node through optimizing the transmit duration in task offloading, CPU frequency in task computing along with the task index to offload in the sequential tasks. In the first case, we decompose the mixed-integer non-convex problem into two levels. In the lower level problem, thanks to the convexity, Karush-Kuhn-Tucker (KKT) conditions are utilized to simplify the problem, which is then solved with bisection search. In the upper level problem, to find solution of the task index to offload, rather than utilizing traversal method, we develop a monotonic condition to simplify the searching process.
In the second case, in order to guarantee the successful computation of the mobile device and cooperative node, the uploading transmission is classified into three schemes. Within each scheme, the non-convex problem is decomposed. In the lower level problem, semi-closed solution is found by Lagrangian dual method. In the upper level problem, traversal method is applied to find the optimal offloading index. 
\end{abstract}

\section{Introduction} \label{s:intro}
Recent years have witnessed an explosive grow of computationally intensive mobile applications such as virtual reality (VR) and augmented reality (AR) \cite{Shi_survey}.
These new applications, usually implemented on mobile devices, provide users with an unprecedented experience, while at the same time pose great challenges to the wireless network by requirement of large scale computation.
On one hand, completing the computation on the mobile device, i.e. local computing, is quite energy consuming and impractical since the mobile device has restrained computation capability.
On the other hand, offloading the computation tasks to the cloud station, which is usually deployed far from the mobile device, brings about long latency in backhaul network transmission and can hardly provide real-time service.
To cope with these challenges, mobile edge computing (MEC) has been proposed as a promising solution \cite{architecture_survey}.
In a MEC system, an edge server rich in computation capability is deployed on the base station (BS) or a network access point.
By offloading the computation tasks to the nearby edge server, the mobile device can complete computation the offloaded data with low latency and save itself from consuming too much energy on local computing \cite{Huang_survey}.

In general, MEC can be divided into two categories by different computation offloading modes: binary offoading \cite{ref_133} and partial offloading \cite{Wu_noma}.
Binary offloading requires the tasks to be either locally computed or offloaded to the edge server as a whole.
In practice, many applications like AR are composed of multiple procedures, making it possible to execute the computation separately, i.e. partial offloading.
In terms of partial offloading, the classical task model is data-partition, in which the input bits of the tasks are independent with each other and thus the tasks can be arbitrarily divided for computation. 

Due to time varying and deep fading, the channel between the mobile device and the BS may not be supportive to offload the computation tasks. 
In this regard, the partial offloading model is more flexible, by investigating the optimal partitioning of the computation tasks considering the channel condition.
Meanwhile, the resource allocation in wireless communication, e.g. transmit power and time, and the capacity of computation central processing unit (CPU), e.g. CPU cores and frequency of the mobile device and the edge server, should be jointly designed, in order for optimal computing performance.
As a pioneering work on single-user MEC, in \cite{ref_81}, the frequency of local CPU cycle and the transmit rate in task offloading are optimized under binary offloading model, for minimizing the energy consumption. Research \cite{ref_83} extend the single-user MEC system into powering with wireless power transfer(WPT) and optimize the power transfer time to maximize energy saving. In addition, \cite{ref_97} investigates partial offloading by jointly optimizing the CPU frequency at the mobile device, the transmit power in data offloading and the ratio of task to be offloaded. 

The design of multi-user MEC system is more complicated, since many mobile devices in the system compete for the communication and computation resources.
To this end, most works in this area optimize the data partition for offloading, transmit power and time for each user, along with the allocation of CPU capacity or virtual machine in the edge server.
As an example of computation resource allocation, \cite{vm} considers the usage of virtual machine in the edge server and investigates a binary multiuser system in which the performance of the edge server degrades with the enlargement of user set.
On the other hand, to develop a optimal offloading strategy for multiple users, \cite{ref_84} studies a partial offloading model working with either time-division multiple access (TDMA) or orthogonal frequency-division multiple access (OFDMA).
In this work, a threshold-based structure is derived based on an offloading priority function, which is related to task characteristics and channel conditions, to decide the data partition to offload and communication resource allocation for each user.
Reference \cite{Wu_noma_delay} extend the communication model into non-orthogonal multiple access (NOMA).
In such a system, the users offload the tasks during the same transmit duration, and their signals are decoded by successive interference cancellation.
In order to minimize the overall delay to finish the tasks, the NOMA transmit duration is optimized along with the offloaded task workload of the users.
Besides, other researches on this topic considers the multi-user system from different aspects.
\cite{Cui_wpt} proposes powering the multiple mobile devices by WPT, and the computation offloading is operated with a TDMA protocal.
\cite{Ding_multiantenna} applies multi-antenna BS to improve the transmission from the mobile device and the BS.
\cite{fd} proposes full-duplex BS and minimizes the sum of energy consumption and task execution latency, with a constraint on interference level.
\cite{ref_87} utilized non-cooperative game theory to investigate the optimal offloading strategy in a code-division multiple access (CDMA) channel, where interference between different channels are taken into account.

The research on multi-user MEC is quite necessary, since there usually exist many mobile devices that can access into the BS (or the network access point) where the edge server is deployed.
As introduced before, most research on multi-user MEC focus on contention scenario, i.e. the users compete for the communication and computation resource in the system.
However, due to the massive number of access and diversity of mobile devices, the mobile devices that are rich in computation capacity or in good channel conditions with the BS can assist the offloading other mobile devices.
For clarification, the former mobile devices is hereinafter referred to as cooperative nodes and the latter mobile devices is referred to as the mobile device.
On one hand, the cooperative nodes can help the mobile devices by providing additional computation resources.
On the other hand, the cooperative nodes in the proximity of both the BS and the mobile device can serve as a relay to improve the transmission.
A few research committed to study this issue, which consider partial offloading in general.
In \cite{ref_127}, a cooperative node utilize its computation resource to help execute the task of the mobile device, and also serve as a relay to help forwarding the task to the BS.
Furthermore, \cite{wpt_coop} considers a cooperative node assists a mobile device to forward the computation task, both of which are powered by WPT.
Apart from assisting the mobile device, the cooperative node has its own task to complete as well.
Finally, to improve the performance of task offloading, \cite{Self_multi_relay} studied the case that a mobile device is aided by multiple cooperative nodes on transmission, under decode-and-forward mode in protocols of TDMA, FDMA and amplify-and-forward mode.

In existing research on cooperative partial offloading MEC systems, only the simple task model of data partition is considered.
In fact, the procedures in complex applications like AR and image recognition are dependent upon each other, making parallel computation less suitable \cite{Huang_survey}.
Take image recognition as an example, which is usually deployed based on neural networks (e.g. convolution neural networks (CNN) and deep neural networks (DNN)) and has layered structure.
The layers, on the other hand, can be viewed as sequential inter-dependent sub-tasks, each of which would be executable only after completing its precedent sub-task.
To finish these sequential inter-dependent sub-tasks, or say sequential tasks, the mostly researched partition based-parallel computation model is no longer suitable.

In this paper, we apply the sequential task model with cooperative mobile edge computing.
In specific, the mobile device has a set of sequential tasks to complete, which is assisted by a cooperative node and an edge server at the BS.
The cooperative node, furthermore, may has its own set of sequential tasks to finish.
By definition of the sequential tasks, there exist $N$ tasks to be executed, which are indexed as task $1$, task $2$, \dots, task $N$. 
The input of task $n$ is the output of task $n-1$.
In this system, our objective is to minimize the overall energy consumption of both the mobile device and the cooperative nodes, while satisfying the time delay requirement for completing the sequential tasks.
To the authors' best knowledge, this is the first work that considers task dependency with the user cooperation in an MEC system.
In this paper, two cases are investigated respectively:
\begin{itemize}
\item The cooperative node has no tasks to complete itself. 
In this case, only the mobile device has a set of sequential tasks to finish.
To complete the computation of sequential tasks the mobile device will first execute the tasks locally and next hand in to the cooperative node through wireless channel.
Upon receiving the computation tasks of the mobile device, the cooperative node, who has no tasks to complete itself, tasks over to execute the tasks with its own computation capacity.
Finally, the cooperative node hand in the computation to the BS through wireless channel, which will complete execution of the sequential tasks.
In this case, an optimization problem is formulated to jointly optimizes the CPU frequencies to execute each task (in the mobile device, the cooperative nodes or in the BS), the transmit durations to hand in the computation and the task indexes to hand in.
The formulated problem is a mix-integer optimization problem, whose optimal solution is hard to find in general.
To overcome this challenge, we first decompose the problem into two levels.
In the lower level, the CPU frequencies and the transmit durations are optimized with the hand in indexes given.
Through steps of mathematical analysis, the related problem is transformed into a single variable problem, which is monotonic within the feasible set.
By proving the activeness of the constraint, the optimal solution can be found by a simple bisection search.
In the upper level, the integer indexes are optimized. In this regard, we derive a monotonic condition of the optimal indexes, to narrow the search of the solution.

\item The cooperative node has tasks to complete itself.
In this case, both the mobile device and the cooperative node has sequential tasks to finish.
Apart from the procedures introduced in the former paragraph, the cooperative node execute computation of its own tasks and hand in to the BS, when the CPU/channel is not occupied by the mobile device.
To avoid resource contention and channel interference, joint scheduling is applied, along with optimization of the CPU frequencies to execute each task (of the mobile evice and the cooperative node), the transmit durations to hand in the computation and the task indexes to hand in.
In specific, the scheduling is divided into three schemes.
For each scheme, an mixed-integer optimizaion problem is formulated, which is decomposed into two levels.
The lower level considering other variables with the given indexes is solved by Lagrangian dual method.
In the upper level, the optimal hand in index is searched.
\end{itemize}

The rest of the paper is organized as follows.
System model is introduced in Section \ref{s:system_model}.
Optimization problems considering two different cases are formulated in Section \ref{s:problem_formulation}.
The solution to the problem in the first case is demonstrated in Section \ref{s:idle_solution}.
The solution to the problem in regards to the second case is presented in Section \ref{s:busy_solution}.
Conclusions are given out in Section \ref{s:conclusion}.
%
%
%
%

\section{System Model} \label{s:system_model}

\begin{figure}
	\begin{center}
		\includegraphics[angle=0,width=0.55 \textwidth]{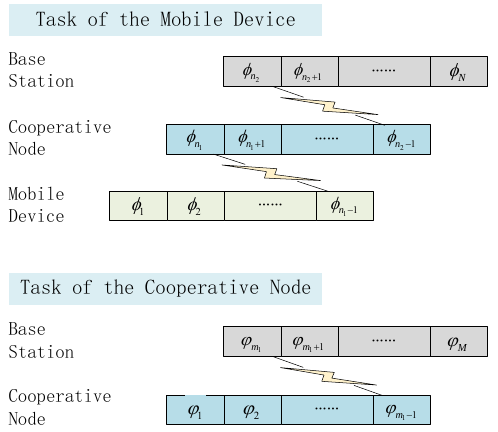}
	\end{center}
	\caption{Task offloading model of the mobile device and the cooperative node.}
	\label{f:system_model}
\end{figure}

Consider a MEC system with one user mobile device, one edge server at the BS, and one cooperative node in the neighborhood of the user mobile device that assists the task offloading and computation.
As shown in Fig.\ref{f:system_model}, the user mobile device has $N$ sequential tasks to complete within given time delay $T_s$, denoted as $\phi_1, \phi_2, \cdots, \phi_N$.
By definition of sequential tasks, the computation of these tasks are related, such that the input of task $\phi_n$ is the output of task $\phi_{n-1}$, for $n=2,\cdots,N$.
Furthermore, task $\phi_n$ is characterized by two parameters $\left(d^s_n,l^s_n\right)$, where $d^s_n$ indicates the input data size of the task (which is in the unit of nats for ease of presentation) and $l^s_n$ indicates the computation capability required to finish the task (which is in the unit of CPU cycles), for $n=1,2,\cdots,N$.

The tasks are generated in the user mobile device. To reduce energy consumption while finishing the computation of the sequential tasks within given time delay, the user mobile device can offload the tasks to the BS, which has abundant computation resource and stable power supply.

In our work, we consider an extreme condition, in which the channel between the user mobile device and the BS is not available for direct transmission due to blockage or deep fading.
Specifically, one cooperative node is considered, to assist the user mobile device on both task transmission and execution.
The cooperative node is a mobile equipment located between the user mobile device and the BS, who also has limited energy, and may have tasks to execute itself.

As shown in Fig.\ref{f:system_model}, note that the cooperative node has $M$ sequential tasks to complete (if the cooperative node has tasks to complete itself) within given time delay $T_r$, which are denoted as $\varphi_1, \varphi_2, \cdots, \varphi_M$. Similarly, task $\varphi_m$ of the cooperative node is characterized by two parameters $\left(d^r_m,l^r_m\right)$, for $m=1,2,\cdots,M$, to indicate the input data size and the required computation capability.

In this system, we consider two cases:
\begin{itemize}
	\item The cooperative node has no tasks to complete itself: In this case, the cooperative node does not have its task to complete.
	The mobile device will first compute the tasks at local, and then handover the task to the cooperative node.
	The cooperative node will in turn compute the tasks and then handover to the BS.
	The BS accomplishs the computation and sends back the results to the mobile device.
	Note that the computation result is usually of small data size and the BS has sufficient power supply and communication resource, the time delay of result feedback can be ignored \cite{ref_97}.
	This case is investigated in Section \ref{s:idle_solution}.
		
	\item The cooperative node has tasks to complete itself: In this case, the cooperative node has its task to complete.
	To guarantee the service of the mobile device, the tasks of the mobile device is of higher priority than that of the cooperative node, i.e. upon receiving the tasks of the mobile device, the cooperative node computes the mobile devices' task prior to its own.
	Furthermore, when the BS has the tasks of the mobile device and of the cooperative node simultaneously, the BS will finish the mobile device's task first.
	Similar to the above case, the mobile device will first compute at local and then handover to the cooperative node.
	The cooperative node, on the other hand, will first compute its own task before the mobile device's task arrives.
	Upon receiving the mobile device's task, the cooperative node will compute the mobile device's task, and continue to compute its own task after handing-over the mobile device's task to the BS.
	Besides, the cooperative node will handover its own task to the BS when the transmit unit of the cooperative node is not occupied.
	Similarly, we ignore the delay for the BS to feedback the computation result to the mobile device and the cooperative node.
	According to this setting, due to the coexistence of uploading process, scheduling of the transmissions are necessarily discussed in Section \ref{s:busy_solution}.
%
%
%
%
	
\end{itemize}

In the following, we introduce the communication and computation model of the mobile device, the cooperative node and the BS.

\subsection{Communication Model}

The required time delay of the tasks are generally a few millisecond. Thus the channel is assumed to be slow fading, and the noise remains unchanged.
For the task of the mobile device, suppose that tasks $\phi_1, \cdots, \phi_{n_1-1}$ are computed at local, tasks $\phi_{n_1}, \cdots, \phi_{n_2-1}$ are computed at the cooperative node, and tasks $\phi_{n_2}, \cdots, \phi_N$ are computed at the BS.
The mobile device should offload the task $\phi_{n_1}$ to the cooperative node, and the cooperative node should offload task $\phi_{n_2}$ to the BS.
The offloading procedures are operated on time slots.
Applying Shannon capacity, we have
\begin{equation} \label{e:mb_1_tran_d}
d^s_{n_1}=
\tau_1B \ln \left(1+\frac{hE^t_1}{\sigma^2 \tau_1}\right),
\end{equation}
\begin{equation} \label{e:mb_2_tran_d}
d^s_{n_2}=\tau_2B \ln \left(1+\frac{gE^t_2}{\sigma^2 \tau_2}\right).
\end{equation}
In (\ref{e:mb_1_tran_d}) and (\ref{e:mb_2_tran_d}), $B$ is the bandwidth allocated to the mobile device and the cooperative node in this system. $h$ is the channel gain from the mobile device to the cooperative node, and $g$ is the channel gain from the cooperative node to the BS..
$E^t_1$ and $\tau_1$ are energy consumption and duration for the mobile device to offload task $\phi_{n_1}$.
$E^t_2$ and $\tau_2$ are energy consumption and duration for the cooperative node to offload task $\phi_{n_2}$.
For the task of the cooperative node, suppose that tasks $\varphi_1, \cdots, \varphi_{m_1-1}$ are computed at the cooperative node and tasks $\varphi_{m_1}, \cdots, \varphi_M$ are computed at the BS, the cooperative node should offload the task $\varphi_{m_1}$ to the BS.
Therefore, we have
\begin{equation} \label{e:rl_tran_d}
d^r_{m_1}=\tau_3 B \ln \left(1+\frac{gE^t_3}{\sigma^2 \tau_3}\right).
\end{equation}
$E^t_3$ and $\tau_3$ are energy consumption and duration for offloading of the cooperative node.
With simple mathematic transformation, (\ref{e:mb_1_tran_d}), (\ref{e:mb_2_tran_d}) and (\ref{e:rl_tran_d}) becomes
\begin{equation} \label{e:md_1_tran_e}
E^t_1=\frac{\sigma^2 \tau_1}{h} \left( e^{\frac{d^s_{n_1}}{\tau_1B}}-1 \right),
\end{equation}
\begin{equation} \label{e:md_2_tran_e}
E^t_2=\frac{\sigma^2 \tau_2}{h} \left( e^{\frac{d^s_{n_2}}{\tau_2B}}-1 \right),
\end{equation}
\begin{equation} \label{e:rl_tran_e}
E^t_3=\frac{\sigma^2 \tau_3}{g} \left( e^{\frac{d^r_{m_1}}{\tau_3B}}-1 \right).
\end{equation}
Note that the system works on given bandwidth which is not separable. To avoid interference, the transmission is carried out on different times slots.
In addition, the value of $h$ and $g$ for can be measured before the task offloading with negligible time overhead.
Note that the mobile device or the cooperative node may not handover the task in the whole process since local computation consumes less energy. For simplicity of exposition, we introduce auxiliary task $\phi^{N+1}$ and $\varphi^{M+1}$ as the exit task, for which the data size and required computation capability are 0. By assuming task task $\phi^{N+1}$ and $\varphi^{M+1}$ are computed at the BS, we have
\begin{equation} \label{e:md_tran_n}
1 \leq n_1 \leq n_2 \leq N+1, n_1 \in \mathcal{Z}, n_2 \in \mathcal{Z},
\end{equation}
and
\begin{equation} \label{e:rl_tran_m}
1 \leq m_1 \leq M+1, m_1 \in \mathcal{Z},
\end{equation}
where $\mathcal{Z}$ is the set of integers.

\subsection{Computation Model}
In terms of computation, it consists of the computation at the mobile device, at the cooperative node and at the BS.
With dynamic voltage scaling of the central processing unit(CPU), the computation capability of the three segments is adjustable.
With this setup, the energy consumption of the computation is expressed by $\kappa_0 f^2$ per cycle, where $\kappa_0$ is the energy coefficient decided by the chip structure, and $f$ denotes CPU frequency.
The time delay of the computation, on the other hand, can be expressed by $l/f$, where $l$ indicates the computation capability required to finish the task.
It is evident in existing research \cite{ref_97} that, to minimize the energy consumption, the CPU frequency remains constant on computing one task. Hence we give out the computation model as follows.

The energy consumption and time delay of the local computation on the mobile device is
\begin{equation} \label{e:md_comp_e}
E^s_n=\kappa_0^s l^s_n {f^s_n}^2, \quad n=1, \cdots, n_1-1,
\end{equation}
\begin{equation} \label{e:md_comp_t}
\tau^s_n=l^s_n/f^s_n, \quad n=1, \cdots, n_1-1.
\end{equation}
In (\ref{e:md_comp_e}) and (\ref{e:md_comp_t}), $\kappa_0^s$ and $\{f^s_n, n< n_1\}$ are energy coefficient and CPU frequency of the mobile device.

As for the cooperative node, the energy consumption and time delay constitute of two parts: computation for the mobile device and for itself. In specific, we have
\begin{equation} \label{e:rl_comp_s_e}
E^{s}_n=\kappa_0^r l^s_n {f^{s}_n}^2, \quad n=n_1, \cdots, n_2-1,
\end{equation}
\begin{equation} \label{e:rl_comp_s_t}
\tau^{s}_n=l^s_n/f^{s}_n, \quad n=n_1, \cdots, n_2-1.
\end{equation}
\begin{equation} \label{e:rl_comp_r_e}
E^r_m=\kappa_0^r l^r_m {f^r_m}^2, \quad m=1, \cdots, m_1-1,
\end{equation}
\begin{equation} \label{e:rl_comp_r_t}
\tau^r_m=l^r_m/f^r_m, \quad m=1, \cdots, m_1-1,
\end{equation}
From (\ref{e:rl_comp_s_e}) to (\ref{e:rl_comp_r_t}), $\kappa_0^r$ is the energy coefficient of the cooperative node. $\{f^{s}_n, n_1 \leq n <n_2\}$ and $\{f^r_m, m \leq m_1\}$ are CPU frequency of the cooperative node to compute task $\phi_n$ and task $\varphi_m$, respectively.

In the BS, the computation capability is larger than the computation capability of the mobile device and the cooperative node.
Upon receiving the tasks from the cooperative node, the BS will compute
Let $\{f^{s}_n, n_2 \leq n \leq N\}$ and $\{f^{r}_m, m_1 \leq m \leq M\}$ be the computation capability allocated to compute task $\phi_n$ and $\varphi_m$ in the BS, the computation delay is expressed by
\begin{equation} \label{e:ed_comp_s_t}
\tau^{s}_n=l^s_n/f^{s}_n, \quad n=n_2, \cdots, N.
\end{equation}
\begin{equation} \label{e:ed_comp_r_t}
\tau^{r}_m=l^r_m/f^{r}_m, \quad m=m_1, \cdots, M.
\end{equation}

Note that $f^{es}(t)$ and $f^{er}(t)$ are related to $\{f^{s}_n, n_2 \leq n \leq N\}$ and $\{f^{r}_m, m_1 \leq m \leq M\}$ such that the former variables indicate the CPU frequency on the BS on a time instant, whereas the latter variables express the CPU frequency on the BS to compute certain tasks. This definition is necessary in solving the case when the cooperative node has its tasks to complete itself. Detailed introduction will be given out in Section \ref{s:busy_solution}. In virtue of stable power supply in the BS, the power consumption of the BS on computing the tasks or sending back the results is no longer taken into account.

\section{Problem Formulation}
\label{s:problem_formulation}
In this section, we formulate the optimization problem based on the system model introduced in Section \ref{s:system_model}.
Our research goal is to reduce the overall energy consumption of the mobile device and the cooperative node, by selecting the proper offloading tasks (i.e. the index of $n_1$, $n_2$ and $m_1$) and adjusting the offloading durations (i.e. $\tau_1$, $\tau_2$ and $\tau_3$) and relative CPU frequencies (i.e. $\{f^{s}_n, 1 \leq n \leq N\}$ and $\{f^{r}_m, 1 \leq m \leq M\}$), while satisfying the delay constraint.
In specific, two problems are established, in which the first one concerns the case when the cooperative node has no task to complete itself, and the second one investigates the case when the cooperative node has its task to complete itself.

When the cooperative node has no tasks to complete, the mobile device will first compute the tasks at local, and then handover the task to the cooperative node.
The cooperative node will in turn compute the tasks and then handover to The BS, where the rest of the tasks are computed.
In this case, the energy consumption of the mobile device consists of (\ref{e:md_1_tran_e}) and the sum of (\ref{e:md_comp_e}), whereas the energy consumption of the cooperative node consists of (\ref{e:md_2_tran_e}) and the sum of (\ref{e:rl_comp_s_e}). The total time delay in computing the tasks is the sum of (\ref{e:md_comp_t}), (\ref{e:rl_comp_s_t}), (\ref{e:ed_comp_s_t}) together with $\tau_1$ and $\tau_2$.
Thereby the optimization is formulated as below.
\begin{prob} \label{p:idle_first}
	\begin{subequations}
		\begin{align}
		\min_{\substack{n_1, n_2,\tau_1, \tau_2\\ \{f^s_n|n=1, \cdots, N\}}}
		&\frac{\sigma^2 \tau_1}{h} \left( e^{\frac{d^s_{n_1}}{\tau_1B}}-1 \right)+ \frac{\sigma^2 \tau_2}{g} \left( e^{\frac{d^s_{n_2}}{\tau_2B}}-1 \right)+ \sum_{n=1}^{n_1-1} \kappa_0^s l^s_n {f^s_n}^2+
		\sum_{n=n_1}^{n_2-1} \kappa_0^r l^s_n {f^{s}_n}^2 \notag \\
		\text{s.t.} \qquad &\sum_{n=1}^{n_1-1} l^s_n/f^s_n + \sum_{n=n_1}^{n_2-1} l^s_n/f^s_n + \sum_{n=n_2}^N l^s_n/f^s_n +\tau_1 +\tau_2 \leq T_s,  \label{e:idle_first_con_time}\\
		&1 \leq n_1 \leq n_2 \leq N+1, \\
		&\tau_1 \geq 0, \tau_2 \geq 0, \\
		&0 \leq f^s_n \leq f^S_{\max}, n=1, 2, \cdots, n_1 \label{e:idle_first_con_f1}\\
		&0 \leq f^s_n \leq f^R_{\max}, n=n_1, \cdots, n_2-1 \label{e:idle_first_con_f2}\\
		&0 \leq f^s_n \leq f^E_{\max}, n=n_2, \cdots, N \label{e:idle_first_con_f3}.
		\end{align}
	\end{subequations}
\end{prob}
In Problem \ref{p:idle_first}, $f^S_{\max}$ and $f^R_{\max}$ are the computation capacity of the mobile device and the BS, respectively.
Furthermore, by denoting
\begin{align}  \label{e:idle_first_con_fall}
f^{\max}_n=\left\{
\begin{array}{ll}
f^S_{\max}, & n=1, 2, \cdots, n_1 \\
f^R_{\max}, & n=n_1, \cdots, n_2-1 \\
f^E_{\max}, & n=n_2, \cdots, N
\end{array} \right\}
\end{align}
the constraints (\ref{e:idle_first_con_f1}), (\ref{e:idle_first_con_f2}) and (\ref{e:idle_first_con_f3}) are merged into $0 \leq f^s_n \leq f^E_{\max}, n=1, 2, \cdots, N$.
Note that (\ref{e:idle_first_con_time}) can be rewritten as $\sum_{n=1}^{N} l^s_n/f^s_n +\tau_1 +\tau_2 \leq T_s$, which will be utilized in the following.

In the case that the cooperative node has tasks to complete, the cooperative node will offload its own task to the BS when the channel is not occupied by tasks of the mobile device.
In this situation, the upload transmission (i.e. for tasks of the mobile device: transmission from the mobile device to the cooperative node, from the cooperative node to the BS and for tasks of the cooperative node: transmission from the cooperative node to the BS) should be scheduled, along with optimization of the variables introduced before.

We consider a general case, in which the mobile device and the cooperative node generate the tasks on different time instant and requires the computation to be finished within different time delay.
Due to similar structure of the problem, we only investigate a certain case that the mobile device generates the tasks first and requires the tasks to be finished before the cooperative node does.
In specific, assume the mobile device and the cooperative node generate the task on the time instant of $0$ and $T_0$, respectively. The mobile device requires the tasks to be finished before time instant $T^S_{th}$ and the cooperative node requires the tasks to be finished before $T^R_{th}$.
In this case, the energy consumption of the mobile device consists of (\ref{e:md_1_tran_e}) and the sum of (\ref{e:md_comp_e}), whereas the energy consumption of the cooperative node consists of (\ref{e:md_2_tran_e}), (\ref{e:rl_tran_e}) and the sum of (\ref{e:rl_comp_s_e}) and (\ref{e:rl_comp_r_e}).
The total time delay for the mobile device to finish the task is the sum of (\ref{e:md_comp_t}), (\ref{e:rl_comp_s_t}), (\ref{e:ed_comp_s_t}) together with $\tau_1$ and $\tau_2$, while the total delay for the cooperative node is the sum of (\ref{e:rl_comp_r_t}), (\ref{e:ed_comp_r_t}) together with $\tau_3$.
Thereby, the problem is formulated as follows.
\begin{prob} \label{p:busy_first}
	\begin{subequations}
		\begin{align}
		\min_{\substack{n_1, n_2, m_1, \tau_1, \tau_2, \tau_3 \\ f^s_s, f^r_s, f^e_s, f^r_r, f^e_r}}
		&\frac{\sigma^2 \tau_1}{h} \left( e^{\frac{d^s_{n_1}}{\tau_1B}}-1 \right)+ \frac{\sigma^2 \tau_2}{g} \left( e^{\frac{d^s_{n_2}}{\tau_2B}}-1 \right) +\frac{\sigma^2 \tau_3}{g} \left( e^{\frac{d^s_{m_1}}{\tau_3B}}-1 \right) \notag \\
		&+\kappa_0^r {f^r_r}^2 \sum_{m=1}^{m_1-1} l^r_m +\kappa_0^s {f^s_s}^2 \sum_{n=1}^{n_1-1} l^s_n+ \kappa_0^r {f^r_s}^2 \sum_{n=n_1}^{n_2-1} l^s_n \notag \\
		\text{s.t.} \qquad
		&\sum_{n=1}^{n_1-1} \frac{l^s_n}{f^s_n} + \sum_{n=n_1}^{n_2-1} \frac{l^s_n}{f^s_n} + \sum_{n=n_2}^N \frac{l^s_n}{f^s_n} +\tau_1 +\tau_2 \leq T^S_{th},  \label{e:busy_identical_f_con_time_s}\\
		&\sum_{m=1}^{m_1-1} \frac{l^r_m}{f^r_m} + \sum_{m=m_1}^{M} \frac{l^r_m}{f^r_m} + \tau_3 \leq T^R_{th} -T_0,  \label{e:busy_identical_f_con_time_r} \\
		&1 \leq n_1 \leq n_2 \leq N+1, \\
		&1 \leq m_1 \leq M+1, \\
		&\tau_1 \geq 0, \tau_2 \geq 0, \tau_3 \geq 0, \\
		&0 \leq f^s_n \leq f^S_{\max}, n=1, 2, \cdots, n_1 \label{e:busy_identical_f_con_f1}\\
		&0 \leq f^s_n \leq f^R_{\max}, n=n_1, \cdots, n_2-1 \label{e:busy_identical_f_con_f2}\\
		&0 \leq f^r_m \leq f^R_{\max}, m=1, 2, \cdots, m_1-1 \label{e:busy_identical_f_con_f3}\\
		&0 \leq f^s_n \leq f^E_{\max}, n=n_2, \cdots, N \label{e:busy_identical_f_con_f4}\\
		&0 \leq f^r_m \leq f^E_{\max}, n=m_1, \cdots, M \label{e:busy_identical_f_con_f5}.
		\end{align}
	\end{subequations}
\end{prob}

\section{Optimal Solution when the cooperative node has no tasks to complete itself} \label{s:idle_solution}
In Problem \ref{p:idle_first}, variables $n_1, n_2, \tau_1, \tau_2, \{f^s_n|n=1, \cdots, N\}$ should be optimized, in which $n_1$ and $n_2$ are integer variables and the rest are continuous variables.
Therefore, Problem \ref{p:idle_first} is a mix-integer optimization problem that is hard to solve.
To find the optimal solution, we decompose Problem \ref{p:idle_first} into two levels.
The lower level problem adjusts $\tau_1, \tau_2, \{f^s_n|n=1, \cdots, N\}$ given a fixed pair of $n_1$ and $n_2$.
The associated optimization problem is
\begin{prob} \label{p:idle_lower}
	\begin{subequations}
		\begin{align}
		U(n_1, n_2) \triangleq \notag \\
		\min_{\substack{\tau_1, \tau_2\\ \{f^s_n|n=1, \cdots, N\}}} &
		\frac{\sigma^2 \tau_1}{h} \left( e^{\frac{d^s_{n_1}}{\tau_1B}}-1 \right)+ \frac{\sigma^2 \tau_2}{g} \left( e^{\frac{d^s_{n_2}}{\tau_2B}}-1 \right)+ \sum_{n=1}^{n_1-1} \kappa_0^s l^s_n {f^s_n}^2+
		\sum_{n=n_1}^{n_2-1} \kappa_0^r l^s_n {f^{s}_n}^2 \notag \\
		\text{s.t.} \qquad &\sum_{n=1}^{N} l^s_n/f^s_n+\tau_1 +\tau_2 \leq T_s,  \label{e:idle_lower_con_time} \\
		&\tau_1 \geq 0, \tau_2 \geq 0, \\
		&0 \leq f^s_n \leq f^{\max}_n, n=n_2, \cdots, N.
		\end{align}
	\end{subequations}
\end{prob}

In the upper level, the following optimization problem finds the optimal $n_1$ and $n_2$, which is equivalent with Problem \ref{p:idle_first}.
\begin{prob} \label{p:idle_upper}
	\begin{subequations}
		\begin{align}
		\min_{n_1, n_2} \quad &U(n_1, n_2) \notag \\
		\text{s.t.} \quad &1 \leq n_1 \leq n_2 \leq N+1.
		\end{align}
	\end{subequations}
\end{prob}

In Problem \ref{p:idle_lower}, the objective function is convex since the second-order derivative of each term is positive.
Furthermore, the function $l^s_n/f^s_n$ in constraint (\ref{e:idle_lower_con_time}) is convex with $f^s_n$.
Hence, Problem \ref{p:idle_lower} is a convex problem and satisfies Slater's condition, hence the Karush-Kuhn-Tucker (KKT) condition is sufficient and necessary condition for the optimal solution \cite{convex_book}.
Before solving Problem \ref{p:idle_lower}, it is easy to claim that
\begin{equation}
f^s_n=f^E_{\max}, n= n_2, \cdots, N.
\end{equation}
Proof of this conclusion is straightforward. The objective function is monotonic decreasing with $\tau_1$ and $\tau_2$. By increasing the $\{f^s_n\}$ for $n= n_2, \cdots, N$, the feasible region of $\tau_1$ and $\tau_2$ are relaxed and the optimal value of the objective function is smaller.


The KKT condition can be listed as follows:
\begin{subequations} \label{e:idle_KKT}
	\small
	\begin{align}
	\lambda + \frac{\sigma^2}{h} \left(e^{\frac{d^s_{n_1}}{B \tau_1}}-1 \right) -\frac{\sigma^2 d^s_{n_1} e^{\frac{d^s_{n_1}}{B \tau_1}}}{Bh\tau_1} -\eta_1 =0 \label{e:idle_KKT_dev_tau1} \\
	\lambda + \frac{\sigma^2}{h} \left(e^{\frac{d^s_{n_2}}{B \tau_2}}-1 \right) -\frac{\sigma^2 d^s_{n_2} e^{\frac{d^s_{n_2}}{B \tau_2}}}{Bh\tau_2} -\eta_2 =0 \label{e:idle_KKT_dev_tau2} \\
	2\kappa_0^s l^s_n f^s_n -\lambda l^s_n /{f^s_n}^2 -\mu_n + \nu_n=0, n=1, 2, \cdots, n_1-1 \label{e:idle_KKT_dev_f1} \\
	2\kappa_0^r l^s_n f^s_n -\lambda l^s_n /{f^s_n}^2 -\mu_n + \nu_n=0, n=n_1, \cdots, n_2-1 \label{e:idle_KKT_dev_f2} \\
	\lambda \left(\sum_{n=1}^N l^s_n/f^s_n +\tau_1 +\tau_2 -T_s \right) =0 \\
	\eta_1 \tau_1 =0, \eta_2 \tau_2 =0 \\
	\mu_n f^s_n=0, n=1, 2, \cdots, n_2-1 \\
	\nu_n \left(f^s_n - f^{\max}_n\right)=0, n=1, 2, \cdots, n_2-1 \\
	\sum_{n=1}^{N} l^s_n/f^s_n+\tau_1 +\tau_2 \leq T_s \label{e:idle_KKT_con_T}\\
	\tau_1 \geq 0, \tau_2 \geq 0 \label{e:idle_KKT_con_t}\\
	0 \leq f^s_n \leq f^{\max}_n, n=1, 2, \cdots, n_2-1 \label{e:idle_KKT_con_f}
	\end{align}
\end{subequations}
in which $\lambda$, $\{\eta_1, \eta_2\}$ and $\{\mu_n, \nu_n\}$ are non-negative Lagrange multipliers associated with constraints (\ref{e:idle_KKT_con_T}), (\ref{e:idle_KKT_con_t}) and (\ref{e:idle_KKT_con_f}), respectively.

With KKT condition in (\ref{e:idle_KKT}), the following theorems can be derived.

\begin{theo} \label{the:idle_optimal_tau}
	The optimal solution of $\tau_1$ and $\tau_2$ in Problem \ref{p:idle_lower} can be expressed as
	\begin{align}
	\tau_1=\frac{d^s_{n_1}}{B \left(W_0 \left(\frac{\lambda h/\sigma^2-1}{e}\right)+1 \right)}, \label{e:idle_optimal_tau1}\\
	\tau_2=\frac{d^s_{n_2}}{B \left(W_0 \left(\frac{\lambda g/\sigma^2-1}{e}\right)+1 \right), \label{e:idle_optimal_tau2}}
	\end{align}
	in which $W_0(x)$ is the principal branch of the Lambert W function, which is defined as the solution of $W_0(x) e^{W_0(x)}=x$ \cite{lambert_w}.
\end{theo}
\begin{IEEEproof}
	In Problem \ref{p:idle_lower}, since $d^s_{n_1}$ and $d^s_{n_2}$ are positive. In order to guarantee successful transmission, $\tau_1$ and $\tau_2$ should be positive as well. Due to similar structure, we will first compute the optimal solution of $\tau_1$.
	In the KKT condition, $\eta_1=0$.
	Let $s_1=\frac{d^s_{n_1}}{B\tau_1}$, equation (\ref{e:idle_KKT_dev_tau1}) can be transformed into
	\begin{equation} \label{e:idle_lemma2_1}
	\left(s_1-1\right)e^{s_1-1}=\frac{\frac{\lambda h}{\sigma^2}-1}{e},
	\end{equation}
	in which $e$ is the base of natural logarithm.
	In (\ref{e:idle_lemma2_1}), by definition of variable $s_1$ and Lagrangian variable $\lambda$, we have $s_1-1>-1$ and the right hand side of the equation is larger than $-\frac{1}{e}$.
	Hence,
	\begin{equation} \label{e:idle_lemma2_2}
	s_1=W_0 \left( \frac{\frac{\lambda h}{\sigma^2}-1}{e} \right) +1,
	\end{equation}
	where $W_0(x)$ is the principal branch of Lambert W function.
	Substitute $s_1=\frac{d^s_{n_1}}{B\tau_1}$ into (\ref{e:idle_lemma2_2}), solution in (\ref{e:idle_optimal_tau1}) is obtained.
	In order to solve $\tau_2$, the process in obtaining (\ref{e:idle_optimal_tau2}) is similar to the above, and omitted for simplification.
	
	This completes the proof.
\end{IEEEproof}

\begin{theo} \label{the:idle_optimal_f}
	The optimal solution of $f^s_n$ in Problem \ref{p:idle_lower} can be expressed as
	\begin{align} \label{e:idle_optimal_fs}
	f^s_n=\left \{
	\begin{array}{ll}
		\min \left \{ {\left(\frac{\lambda}{2\kappa_0^s}\right)}^{\frac{1}{3}}, f^S_{\max} \right \}, & n=1, 2, \cdots, n_1-1 \\
		\min \left \{ {\left(\frac{\lambda}{2\kappa_0^r}\right)}^{\frac{1}{3}}, f^R_{\max} \right \}, & n=n_1, \cdots, n_2-1
	\end{array} \right \}
	\end{align}
 \end{theo}
\begin{IEEEproof}
	At first, it is noted that $f^s_n \neq 0, \forall n=1, 2, \cdots, n_2-1$ due to constraint (\ref{e:idle_KKT_con_T}). Hence $\mu_n=0$. Resorting to (\ref{e:idle_first_con_fall}), the set $n=1, 2, \cdots, n_1-1$ can be divided into two subsets: $\mathcal{N}_1=\{n|0<f^s_n<f^S_{\max}\}$ and $\mathcal{N}_2=\{n|f^s_n=f^S_{\max}\}$. The set $n=n_1, \cdots, n_2-1$ can be divided into two subsets: $\mathcal{N}_3=\{n|0<f^s_n<f^R_{\max}\}$ and $\mathcal{N}_4=\{n|f^s_n=f^R_{\max}\}$.
	
	For $n\in \mathcal{N}_1$, $\nu_n=0$. Therefore, equation (\ref{e:idle_KKT_dev_f1}) can be transformed into
	\begin{equation}
	2\kappa_0^s l^s_n f^s_n=\lambda \frac{l^s_n}{{f^s_n}^2}, \forall n \in \mathcal{N}_1,
	\end{equation}
	which further indicates that
	\begin{equation} \label{e:idle_lemma1_1}
	f^s_n={\left(\frac{\lambda}{2\kappa_0^s}\right)}^{\frac{1}{3}}, \forall n \in \mathcal{N}_1.
	\end{equation}
	
	For $n\in \mathcal{N}_2$, equation (\ref{e:idle_KKT_dev_f1}) turns into
	\begin{equation} \label{e:idle_lemma1_2}
	2\kappa_0^s l^s_n f^s_n +\nu_n =\lambda \frac{l^s_n}{{f^s_n}^2}, \forall n \in \mathcal{N}_2.
	\end{equation}
	Substitute $f^s_n=f^S_{\max}$ into (\ref{e:idle_lemma1_2}) gets to
	\begin{equation} \label{e:idle_lemma1_3}
	\lambda=2\kappa_0^s {f^S_{\max}}^3 +\nu_n \frac{{f^S_{\max}}^2}{l^s_n}, \forall n \in \mathcal{N}_2.
	\end{equation}
	Suppose the set $\mathcal{N}_1$ and $\mathcal{N}_2$ are both nonempty. Combine the equation (\ref{e:idle_lemma1_1}) and (\ref{e:idle_lemma1_3}), we have
	\begin{align}
	\lambda&=2\kappa_0^s {f^s_n}^3, &\forall n\in \mathcal{N}_1, \label{e:idle_lemma1_5}\\
	\lambda&=2\kappa_0^s {f^S_{\max}}^3 +\nu_n \frac{{f^S_{\max}}^2}{l^s_n}, &\forall n\in \mathcal{N}_2 \label{e:idle_lemma1_6}.
	\end{align}
	Since $f^s_n<f^S_{\max}$ and $\nu_n \geq 0$, (\ref{e:idle_lemma1_5}) contradicts with (\ref{e:idle_lemma1_6}). Thus either $\mathcal{N}_1=\emptyset$ or $\mathcal{N}_2=\emptyset$.
	In other words, we have
	\begin{equation} \label{e:idle_lemma1_7}
	f^s_n={\left(\frac{\lambda}{2\kappa_0^s}\right)}^{\frac{1}{3}}, n=1, 2, \cdots, n_1-1,
	\end{equation}
	or
	\begin{equation} \label{e:idle_lemma1_8}
	f^s_n=f^S_{\max}, n=1, 2, \cdots, n_1-1.
	\end{equation}
	Following the same reasoning, either $\mathcal{N}_3=\emptyset$ or $\mathcal{N}_4=\emptyset$, i.e. we have
	\begin{equation} \label{e:idle_lemma1_9}
	f^s_n={\left(\frac{\lambda}{2\kappa_0^r}\right)}^{\frac{1}{3}}, n=n_1, \cdots, n_2-1,
	\end{equation}
	or
	\begin{equation} \label{e:idle_lemma1_10}
	f^s_n=f^R_{\max}, n=n_1, \cdots, n_2-1.
	\end{equation}
	
	This completes the proof.
\end{IEEEproof}

So far, the optimal solution of Problem \ref{p:idle_lower} is expressed with single Lagrangian variable $\lambda$.
To minimize the objective function of Problem \ref{p:idle_lower}, the optimal solution of $\tau_1$ and $\tau_2$ should be as large as possible, which causes the activeness of constraint (\ref{e:idle_lower_con_time}).
\begin{theo} \label{the:idle_optimal_monotonic}
	The left hand side of constraint (\ref{e:idle_lower_con_time}) is monotonically decreasing with $\lambda$.
\end{theo}
\begin{IEEEproof}
	The left hand side of (\ref{e:idle_lower_con_time}) constitutes of three terms.
	According to Theorem \ref{the:idle_optimal_tau}, since the principal branch of Lambert W function, i.e. $W_0(x)$ is increasing, the optimal solution of $\tau_1$ and $\tau_2$ are monotonic decreasing with respect to $\lambda$.
	In reference to Theorem \ref{the:idle_optimal_f}, $f^s_n$, $n=1, 2, \cdots, n_2-1$ are non-decreasing functions of $\lambda$. $f^s_n=f^E_{\max}$, $n=n_2, \cdots, N$, which are constants. Thus the term $\sum_{n=1}^{N} l^s_n/f^s_n$ is non-increasing with respect to $\lambda$.
	
	This completes the proof.
\end{IEEEproof}
Thanks to the monotonic increasing property in Theorem \ref{the:idle_optimal_monotonic}, the optimal solution of Problem \ref{p:idle_lower} can be found by bisection search with respect to the Lagrangian variable $\lambda$.

Next we turn to solve Problem \ref{p:idle_upper}. Since the index $n_1$ and $n_2$ are integers, traversal can be utilized to find the optimal solution.
To further reduce the computation complexity, the following theorem is expected.
\begin{theo} \label{the:idle_d_decrease}
	For the cooperative node, the data size of the offloaded task $\phi_{n_2}$ has the following characteristics.
	If
	\begin{equation}
	U(n_1,n_2) < U(n_1,n_2-1),
	\end{equation}
	then
	\begin{equation}
	d^s_{n_2} < d^s_{n_2-1}.
	\end{equation}
\end{theo}
\begin{IEEEproof}
	Suppose $d^s_{n_2} \geq d^s_{n_2-1}$, $\lambda^{\dagger}$ and $\lambda^{\ddagger}$ are optimal Lagrangian multipliers for $U(n_1,n_2)$ and $U(n_1,n_2-1)$, respectively.
	We prove Theorem \ref{the:idle_d_decrease} by contradiction.
	For purpose of simplification, we define functions $f_1(x)= B\left(W_0 \left(\frac{x h/\sigma^2-1}{e}\right)+1 \right)$,
	$f_2(x)= B\left(W_0 \left(\frac{x g/\sigma^2-1}{e}\right)+1 \right)$,
	$f_3(x)=\min \left \{ {\left(\frac{x}{2\kappa_0^s}\right)}^{\frac{1}{3}}, f^S_{\max} \right \}$ and
	$f_4(x)=\min \left \{ {\left(\frac{x}{2\kappa_0^r}\right)}^{\frac{1}{3}}, f^R_{\max} \right \}$.
	Note that $f_1(x)$ and $f_2(x)$ are increasing functions, and $f_3(x)$ and $f_4(x)$ are non-decreasing functions.
	Denote $\{\tau_1^{\dagger}, \tau_2^{\dagger}, \{f_n^{s,\ddagger} \}\}$ and $\{\tau_1^{\ddagger}, \tau_2^{\ddagger}, \{f_n^{s,\ddagger} \}\}$ as optimal solutions of the lower level problem relative to $U(n_1,n_2)$ and $U(n_1,n_2-1)$.
	By utilizing Theorem \ref{the:idle_optimal_tau}, Theorem \ref{the:idle_optimal_f} and the activeness of constraint (\ref{e:idle_lower_con_time}), the following equations are established.
	\begin{align}
	\sum_{n=1}^{n_1-1} \frac{l^s_{n}}{f_3(\lambda^{\dagger})}+
	\sum_{n=n_1}^{n_2-1} \frac{l^s_{n}}{f_4(\lambda^{\dagger})}+
	\sum_{n=n_2}^{N} \frac{l^s_{n}}{f^E_{\max}}  +\frac{d^s_{n_1}}{f_1(\lambda^{\dagger})}+
	\frac{d^s_{n_2}}{f_2(\lambda^{\dagger})}&= T_s, \label{e:idle_d_decrease_1}\\
	\sum_{n=1}^{n_1-1} \frac{l^s_{n}}{f_3(\lambda^{\ddagger})}+
	\sum_{n=n_1}^{n_2-2} \frac{l^s_{n}}{f_4(\lambda^{\ddagger})}+
	\sum_{n=n_2-1}^{N} \frac{l^s_{n}}{f^E_{\max}}  +\frac{d^s_{n_1}}{f_1(\lambda^{\ddagger})}+
	\frac{d^s_{n_2-1}}{f_2(\lambda^{\ddagger})}&= T_s. \label{e:idle_d_decrease_2}
	\end{align}
	Substract (\ref{e:idle_d_decrease_1}) with (\ref{e:idle_d_decrease_2}), we have
	\begin{align} \label{e:idle_d_decrease_3}
	\sum_{n=1}^{n_1-1} \frac{l^s_{n}}{f_3(\lambda^{\dagger})}+
	\sum_{n=n_1}^{n_2-2} \frac{l^s_{n}}{f_4(\lambda^{\dagger})}+ \frac{l^s_{n_2-1}}{f_4(\lambda^{\dagger})}+
	\frac{d^s_{n_1}}{f_1(\lambda^{\dagger})}+
	\frac{d^s_{n_2}}{f_2(\lambda^{\dagger})} \notag \\ =
	\sum_{n=1}^{n_1-1} \frac{l^s_{n}}{f_3(\lambda^{\ddagger})}+
	\sum_{n=n_1}^{n_2-2} \frac{l^s_{n}}{f_4(\lambda^{\ddagger})}+
	\frac{l^s_{n_2-1}}{f^E_{\max}}+
	\frac{d^s_{n_1}}{f_1(\lambda^{\ddagger})}+
	\frac{d^s_{n_2-1}}{f_2(\lambda^{\ddagger})}.
	\end{align}
	Compare the left hand side and the right hand side of equation (\ref{e:idle_d_decrease_3}), the following inequality is expected.
	\begin{align} \label{e:idle_d_decrease_4}
	\sum_{n=1}^{n_1-1} \frac{l^s_{n}}{f_3(\lambda^{\dagger})}+
	\sum_{n=n_1}^{n_2-2} \frac{l^s_{n}}{f_4(\lambda^{\dagger})}+
	\frac{d^s_{n_1}}{f_1(\lambda^{\dagger})}+
	\frac{d^s_{n_2-1}}{f_2(\lambda^{\dagger})} \notag \\ \leq
	\sum_{n=1}^{n_1-1} \frac{l^s_{n}}{f_3(\lambda^{\dagger})}+
	\sum_{n=n_1}^{n_2-2} \frac{l^s_{n}}{f_4(\lambda^{\dagger})}+
	\frac{d^s_{n_1}}{f_1(\lambda^{\dagger})}+
	\frac{d^s_{n_2}}{f_2(\lambda^{\dagger})} \notag \\ \leq
	\sum_{n=1}^{n_1-1} \frac{l^s_{n}}{f_3(\lambda^{\ddagger})}+
	\sum_{n=n_1}^{n_2-2} \frac{l^s_{n}}{f_4(\lambda^{\ddagger})}+
	\frac{d^s_{n_1}}{f_1(\lambda^{\ddagger})}+
	\frac{d^s_{n_2-1}}{f_2(\lambda^{\ddagger})},
	\end{align}
	in which the first inequality is derived on the basis of $d^s_{n_2} \geq d^s_{n_2-1}$, and the second inequality is due to $f_4(\lambda^{\dagger}) \leq f^E_{\max}$.
	Utilizing (\ref{e:idle_d_decrease_4}), it can be inferred that $\lambda^{\dagger} \geq \lambda^{\ddagger}$.
	Resorting to (\ref{e:idle_optimal_tau1}), (\ref{e:idle_optimal_tau2}) and (\ref{e:idle_optimal_fs}),
	\begin{align}
		&\tau_1^{\dagger} \leq \tau_1^{\ddagger}, \label{e:idle_d_decrease_5}\\
		&\tau_2^{\dagger} \leq \tau_2^{\ddagger}, \label{e:idle_d_decrease_6}\\
		&f_n^{s,\dagger} \geq f_n^{s,\ddagger}, n=1, 2, \cdots, N. \label{e:idle_d_decrease_7}
	\end{align}
	Substitute $\{\tau_1^{\dagger}, \tau_2^{\dagger}, \{f_n^{s,\ddagger} \}\}$ and $\{\tau_1^{\ddagger}, \tau_2^{\ddagger}, \{f_n^{s,\ddagger} \}\}$ into the objective function of Problem \ref{p:idle_lower} respectively, it can be checked that $U(n_1, n_2) \geq U(n_1, n_2-1)$.
	
	This completes the proof.
\end{IEEEproof}

\begin{rem} \label{rem:idle_cpu_frequency}
	Theorem \ref{the:idle_optimal_f} demonstrates that the optimal CPU frequency to solve a sequence of tasks in certain site is identical.
	Particularly, for sequential tasks of the mobile device, CPU frequency for computing at local, at the cooperative node and at the BS, respectively, are identical.
\end{rem}

\begin{rem} \label{rem:idle_d_decrease}
	Theorem \ref{the:idle_d_decrease} provide additional insight on the solution of index $n_2$. In specific, the optimal $n_2$ should satisfy $d_{n_2} \leq d_{n_2-1}$. This narrow the search of the optimal solution.
\end{rem}

\section{Optimal Solution when the cooperative node has tasks to complete itself} \label{s:busy_solution}
In this section, Problem \ref{p:busy_first} that considers the case when the cooperative node and the mobile device both has sequential tasks to compute is solved.
In Section \ref{s:idle_solution}, Theorem \ref{the:idle_optimal_f} has demonstrated that CPU frequency for computing the sequential tasks of the mobile device at local, at the cooperative node and at the BS are identical, respectively.
As for tasks of the cooperative node, similar to the former conclusion, the CPU frequency for computing at the cooperative node and at the BS are identical \footnote{The proof of this conclusion is similar to the proof of Theorem \ref{the:idle_optimal_f} and thus neglected for simplification.}.

In this system, we denote $f^s_s$ as the CPU frequency of the mobile device to compute its own task, $f^r_s$ and $f^r_r$ as the CPU frequency of the cooperative node to compute the tasks for the mobile device and the cooperative node, $f^e_s$ and $f^e_r$ as the CPU frequency of the BS to compute the tasks for the mobile device and the cooperative node.
Due to the priority of the mobile device's task, we define the time for computation of the mobile device's task on the BS is $\tau_s$.

In this case, the communication resource need to be scheduled to guarantee successful transmission from the mobile device to the cooperative node and from the cooperative node to the BS.
Define $T_1$, $T_2$ and $T_3$ as the time delay for the computation of the mobile device 's task at local, computation of the mobile device's task at the cooperative node and computation of the cooperative node 's task at the cooperative node, respectively.
In other words,
\begin{align}
&T_1 = \frac{\sum_{n=1}^{n_1-1} l^s_n}{f^s_s} \\
&T_2 = \frac{\sum_{n=n_1}^{n_2-1} l^s_n}{f^r_s} \\
&T_3 = \frac{\sum_{m=1}^{m_1-1} l^r_m}{f^r_r}
\end{align}

\begin{figure}
	\begin{center}
		\includegraphics[angle=0,width=0.70 \textwidth]{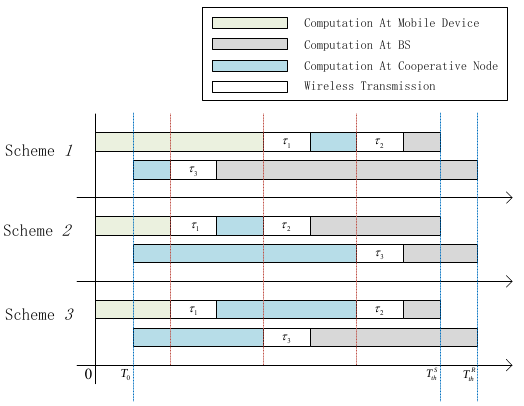}
	\end{center}
	\caption{Three different Schemes when the cooperative node has tasks to complete itself.}
	\label{f:time_schemes}
\end{figure}

As shown in Fig. \ref{f:time_schemes}, the uploading transmission is classified into three schemes, which are
\begin{itemize}
	\item Scheme 1: $\tau_3 \rightarrow \tau_1 \rightarrow \tau_2$,
	\item Scheme 2: $\tau_1 \rightarrow \tau_2 \rightarrow \tau_3$,
	\item Scheme 3: $\tau_1 \rightarrow \tau_3 \rightarrow \tau_2$.
\end{itemize}
In Scheme 1, Problem \ref{p:busy_first} can be transformed into the following form \footnote{The optimization problem in Scheme 2 and Scheme 3 are in similar structure with Scheme 1, and the problem solution follows the same procedure. In this version, optimization problems in Scheme 2 and Scheme 3 are given out in Appendix \ref{s:appendix_problems}, whereas the solution for these problems are left out.}.
\begin{prob} \label{p:busy_case_1}
	\begin{subequations}
		\begin{align}
		\min_{\substack{n_1, n_2, m_1, \tau_1, \tau_2, \tau_3 \\ T_1, T_2, T_3, \tau_s}}
		&\frac{\sigma^2 \tau_1}{h} \left( e^{\frac{d^s_{n_1}}{\tau_1B}}-1 \right)+ \frac{\sigma^2 \tau_2}{g} \left( e^{\frac{d^s_{n_2}}{\tau_2B}}-1 \right) +\frac{\sigma^2 \tau_3}{g} \left( e^{\frac{d^s_{m_1}}{\tau_3B}}-1 \right) \notag \\
		&+\frac{\kappa_0^r \left( \sum_{m=1}^{m_1-1} l^r_m \right)^3}{T_3^2} +\frac{\kappa_0^s \left( \sum_{n=1}^{n_1-1} l^s_n \right)^3}{T_1^2} + \frac{\kappa_0^r \left( \sum_{n=n_1}^{n_2-1} l^s_n \right)^3}{T_2^2} \notag \\
		\text{s.t.} \qquad &T_0+T_3+ \tau_3 \leq T_1, \label{e:busy_case_1_con_time} \\
		&\frac{\sum_{n=1}^{n_1-1} l^s_n}{T_1} \leq f^S_{\max}, \label{e:busy_case_1_con_f1} \\
		&\frac{\sum_{n=n_1}^{n_2-1} l^s_n}{T_2} \leq f^R_{\max}, \label{e:busy_case_1_con_f2} \\	
		&\frac{\sum_{m=1}^{m_1-1} l^r_n}{T_3} \leq f^R_{\max}, \label{e:busy_case_1_con_f3} \\
		&\sum_{n=n_2}^{N} l^s_n \leq f^E_{\max} \tau_s, \label{e:busy_case_1_con_mb} \\
		&\sum_{m=m_1}^{M} l^r_m \leq f^E_{\max} \left(T^R_{th} -T_0 -T_3 -\tau_3 -\tau_s \right), \label{e:busy_case_1_con_rl} \\
		&\tau_s \leq T^S_{th} -T_1 -\tau_1 - T_2 -\tau_2. \label{e:busy_case_1_con_taus}
		\end{align}
	\end{subequations}
\end{prob}
In Problem \ref{p:busy_case_1}, equation (\ref{e:busy_case_1_con_time}) is the restriction of scheme 1. (\ref{e:busy_case_1_con_f1}), (\ref{e:busy_case_1_con_f2}) and (\ref{e:busy_case_1_con_f3}) are constraint of the computation capability on the mobile device and the cooperative node. (\ref{e:busy_case_1_con_mb}) and (\ref{e:busy_case_1_con_rl}) are computation capability constraint of the BS, in order to compute the tasks of the mobile device and the cooperative node, respectively. Finally, (\ref{e:busy_case_1_con_taus}) guarantees that the tasks offloaded to the BS are computed before the required deadline.

\begin{assu} \label{assu:busy_problem_simplification}
	In Problem \ref{p:busy_case_1}, major challenge in solution of the problem lies in the allocation of slots on wireless transmission and task computation.
	To simplify the problem and further investigate the analytical structure of Problem \ref{p:busy_case_1}, we assume that constraints (\ref{e:busy_case_1_con_f1}), (\ref{e:busy_case_1_con_f2}) and (\ref{e:busy_case_1_con_f3}) always satisfies, i.e. these constraints are no longer taken into account.
\end{assu}

Next we turn to solve Problem \ref{p:busy_case_1}. Problem \ref{p:busy_case_1} is a mix-integer problem which is hard to solve with traditional methods.
To find optimal solution, we decompose Problem \ref{p:busy_case_1} into two levels. The lower level problem adjusts $\tau_1, \tau_2, \tau_3, T_1, T_2, T_3, \tau_s$ given fixed $n_1$, $n_2$ and $m_1$, whereas the upper level problem search for the optimal $n_1$, $n_2$ and $m_1$.

In the lower level, the associated optimization problem is
\begin{prob} \label{p:busy_case_1_lower}
	\begin{subequations}
		\begin{align}
		U(n_1, n_2, m_1) \triangleq &\notag \\
		\min_{\substack{\tau_1, \tau_2, \tau_3 \\ T_1, T_2, T_3, \tau_s}} \qquad
		&\frac{\sigma^2 \tau_1}{h} \left( e^{\frac{d^s_{n_1}}{\tau_1B}}-1 \right)+ \frac{\sigma^2 \tau_2}{g} \left( e^{\frac{d^s_{n_2}}{\tau_2B}}-1 \right) +\frac{\sigma^2 \tau_3}{g} \left( e^{\frac{d^s_{m_1}}{\tau_3B}}-1 \right) \notag \\
		&+\frac{\kappa_0^r \left( \sum_{m=1}^{m_1-1} l^r_m \right)^3}{T_3^2} +\frac{\kappa_0^s \left( \sum_{n=1}^{n_1-1} l^s_n \right)^3}{T_1^2} + \frac{\kappa_0^r \left( \sum_{n=n_1}^{n_2-1} l^s_n \right)^3}{T_2^2} \notag \\
		\text{s.t.} \qquad &T_0+T_3+ \tau_3 \leq T_1, \label{e:busy_case_1_lower_con_time} \\
		&\frac{\sum_{n=1}^{n_1-1} l^s_n}{T_1} \leq f^S_{\max}, \label{e:busy_case_1_lower_con_f1} \\
		&\frac{\sum_{n=n_1}^{n_2-1} l^s_n}{T_2} \leq f^R_{\max}, \label{e:busy_case_1_lower_con_f2} \\	
		&\frac{\sum_{m=1}^{m_1-1} l^r_n}{T_3} \leq f^R_{\max}, \label{e:busy_case_1_lower_con_f3} \\
		&\sum_{n=n_2}^{N} l^s_n \leq f^E_{\max} \tau_s, \label{e:busy_case_1_lower_con_mb} \\
		&\sum_{m=m_1}^{M} l^r_m \leq f^E_{\max} \left(T^R_{th} -T_0 -T_3 -\tau_3 -\tau_s \right), \label{e:busy_case_1_lower_con_rl} \\
		&\tau_s \leq T^S_{th} -T_1 -\tau_1 - T_2 -\tau_2. \label{e:busy_case_1_lower_con_taus}
		\end{align}
	\end{subequations}
\end{prob}

In the upper level, the following optimization problem finds the optimal $n_1$, $n_2$ and $m_1$, which is equivalent with Problem \ref{p:busy_case_1}.
\begin{prob} \label{p:busy_case_1_upper}
	\begin{subequations}
		\begin{align}
		\min_{n_1, n_2} \quad &U(n_1, n_2, m_1) \notag \\
		\text{s.t.} \quad &1 \leq n_1 \leq n_2 \leq N+1, \\
		&1 \leq m_1 \leq M+1.
		\end{align}
	\end{subequations}
\end{prob}

Next, we solve Problem \ref{p:busy_case_1_lower} in the lower level first.
Before going into solution of Problem \ref{p:busy_case_1_lower}, the following theorem can be derived.
\begin{theo} \label{the:busy_optimal_taus}
	The optimal function value of Problem \ref{p:busy_case_1_lower} is irrelevant to the value of $\tau_s$, yet constraint (\ref{e:busy_case_1_lower_con_mb}), (\ref{e:busy_case_1_lower_con_rl}) and (\ref{e:busy_case_1_lower_con_taus}) are equivalent to the following constraints
	\begin{align} \label{e:busy_case_1_lower_transform}
	\frac{\sum_{n=n_2}^{N} l^s_n }{f^E_{\max}} \leq \min \Big \{T^R_{th} -T_0 -T_3 -\tau_3 - \frac{\sum_{m=m_1}^{M} l^r_m}{f^E_{\max}}, T^S_{th} -T_1 -\tau_1 - T_2 -\tau_2 \Big \}
	\end{align}
\end{theo}
\begin{IEEEproof}
	The objective function of Problem \ref{p:busy_case_1_lower} is function of $\tau_1$, $\tau_2$, $\tau_3$, $T_1$, $T_2$ and $T_3$, the solution of $\tau_s$ can be transformed into a feasibility problem
	\begin{prob} \label{p:busy_case_1_taus}
		\begin{subequations}
			\begin{align}
			\text{find} \quad & \quad \tau_s \notag\\
			\text{s.t.} \quad & \frac{\sum_{n=n_2}^{N} l^s_n }{f^E_{\max}} \leq \tau_s, \label{e:busy_case_1_taus_con_mb} \\
			&\tau_s \leq T^R_{th} -T_0 -T_3 -\tau_3 - \frac{\sum_{m=m_1}^{M} l^r_m}{f^E_{\max}}, \label{e:busy_case_1_taus_con_rl} \\
			&\tau_s \leq T^S_{th} -T_1 -\tau_1 - T_2 -\tau_2. \label{e:busy_case_1_taus_con_taus}
			\end{align}
		\end{subequations}
	\end{prob}
	In Problem \ref{p:busy_case_1_taus}, the constraint (\ref{e:busy_case_1_taus_con_mb}), (\ref{e:busy_case_1_taus_con_rl}) and (\ref{e:busy_case_1_taus_con_taus}) are derived from (\ref{e:busy_case_1_lower_con_mb}), (\ref{e:busy_case_1_lower_con_rl}) and (\ref{e:busy_case_1_lower_con_taus}) in Problem \ref{p:busy_case_1_lower}, respectively.
	Since (\ref{e:busy_case_1_taus_con_mb}), (\ref{e:busy_case_1_taus_con_rl}) and (\ref{e:busy_case_1_taus_con_taus}) are linear constraints, it is straightforward that Theorem \ref{the:busy_optimal_taus} establishes.
\end{IEEEproof}

Utilize Theorem \ref{the:busy_optimal_taus}, Problem \ref{p:busy_case_1_lower} can be transformed into the following form.
\begin{prob} \label{p:busy_case_1_lower_transform}
	\begin{subequations}
		\begin{align}
		U(n_1, n_2, m_1) \triangleq &\notag \\
		\min_{\substack{\tau_1, \tau_2, \tau_3 \\ T_1, T_2, T_3, \tau_s}} \qquad
		&\frac{\sigma^2 \tau_1}{h} \left( e^{\frac{d^s_{n_1}}{\tau_1B}}-1 \right)+ \frac{\sigma^2 \tau_2}{g} \left( e^{\frac{d^s_{n_2}}{\tau_2B}}-1 \right) +\frac{\sigma^2 \tau_3}{g} \left( e^{\frac{d^s_{m_1}}{\tau_3B}}-1 \right) \notag \\
		&+\frac{\kappa_0^r \left( \sum_{m=1}^{m_1-1} l^r_m \right)^3}{T_3^2} +\frac{\kappa_0^s \left( \sum_{n=1}^{n_1-1} l^s_n \right)^3}{T_1^2} + \frac{\kappa_0^r \left( \sum_{n=n_1}^{n_2-1} l^s_n \right)^3}{T_2^2} \notag \\
		\text{s.t.} \qquad &T_0+T_3+ \tau_3 \leq T_1, \label{e:busy_case_1_lower_transform_con_time} \\
		&\frac{\sum_{n=1}^{n_1-1} l^s_n}{f^S_{\max}} \leq T_1, \label{e:busy_case_1_lower_transform_con_f1} \\
		&\frac{\sum_{n=n_1}^{n_2-1} l^s_n}{f^R_{\max}} \leq T_2, \label{e:busy_case_1_lower_transform_con_f2} \\	
		&\frac{\sum_{m=1}^{m_1-1} l^r_n}{f^R_{\max}} \leq T_3, \label{e:busy_case_1_lower_transform_con_f3} \\
		&\frac{\sum_{n=n_2}^{N} l^s_n }{f^E_{\max}} +\frac{\sum_{m=m_1}^{M} l^r_m}{f^E_{\max}} +T_0 +T_3 +\tau_3 \leq T^R_{th}, \label{e:busy_case_1_lower_transform_con_taus1} \\
		&\frac{\sum_{n=n_2}^{N} l^s_n }{f^E_{\max}} +T_1 +\tau_1 +T_2 +\tau_2\leq T^S_{th}. \label{e:busy_case_1_lower_transform_con_taus2}
		\end{align}
	\end{subequations}
\end{prob}
In Problem \ref{p:busy_case_1_lower_transform}, constraints (\ref{e:busy_case_1_lower_transform_con_taus1}) and (\ref{e:busy_case_1_lower_transform_con_taus2}) are from (\ref{e:busy_case_1_lower_transform}).

Problem \ref{p:busy_case_1_lower_transform} is a convex problem since the objective function is convex and the constraints are linear, which can be solved by standard methods such as interior point method.
Alternatively, to gain more essential insights, Lagrangian duality method used to obtain a structure of optimal solution.
Let $\lambda$, $\eta_1$ and $\eta_2$ be the positive dual variables associated with (\ref{e:busy_case_1_lower_transform_con_time}), (\ref{e:busy_case_1_lower_transform_con_taus1}) and (\ref{e:busy_case_1_lower_transform_con_taus2}), respectively.
Thereby, the partial Lagrangian of Problem \ref{p:busy_case_1_lower_transform} is given by
\begin{align} \label{e:busy_case_1_lagrangian}
& \mathcal{L} (\tau_1, \tau_2, \tau_3, T_1, T_2, T_3, \lambda, \eta_1, \eta_2) \notag \\
= & \frac{\sigma^2 \tau_1}{h} \left( e^{\frac{d^s_{n_1}}{\tau_1B}}-1 \right)+ \frac{\sigma^2 \tau_2}{g} \left( e^{\frac{d^s_{n_2}}{\tau_2B}}-1 \right) +\frac{\sigma^2 \tau_3}{g} \left( e^{\frac{d^s_{m_1}}{\tau_3B}}-1 \right) \notag \\
& +\frac{\kappa_0^r \left( \sum_{m=1}^{m_1-1} l^r_m \right)^3}{T_3^2} +\frac{\kappa_0^s \left( \sum_{n=1}^{n_1-1} l^s_n \right)^3}{T_1^2} + \frac{\kappa_0^r \left( \sum_{n=n_1}^{n_2-1} l^s_n \right)^3}{T_2^2} \notag \\
& +\lambda \left( T_0+T_3+ \tau_3 - T_1 \right) \notag \\
& + \eta_1 \left( \frac{\sum_{n=n_2}^{N} l^s_n }{f^E_{\max}} +\frac{\sum_{m=m_1}^{M} l^r_m}{f^E_{\max}} +T_0 +T_3 +\tau_3 - T^R_{th} \right) \notag \\ 
& + \eta_2 \left( \frac{\sum_{n=n_2}^{N} l^s_n }{f^E_{\max}} +T_1 +\tau_1 +T_2 +\tau_2 - T^S_{th} \right).
\end{align}
Then the dual function of Problem \ref{p:busy_case_1_lower_transform} is given by $\mathcal{G} (\lambda, \eta_1, \eta_2)$, which is defined in Problem \ref{p:busy_case_1_lagrangian}.
\begin{prob} \label{p:busy_case_1_lagrangian}
	\begin{subequations}
		\begin{align}
		\mathcal{G} (\lambda, \eta_1, \eta_2) = \min_{\substack{\tau_1, \tau_2, \tau_3 \\ T_1, T_2, T_3}} & \mathcal{L} (\tau_1, \tau_2, \tau_3, T_1, T_2, T_3, \lambda, \eta_1, \eta_2) \notag \\
		\text{s.t.} \quad &\frac{\sum_{n=1}^{n_1-1} l^s_n}{f^S_{\max}} \leq T_1, \label{e:busy_case_1_lagrangian_con_f1} \\
		&\frac{\sum_{n=n_1}^{n_2-1} l^s_n}{f^R_{\max}} \leq T_2, \label{e:busy_case_1_lagrangian_con_f2} \\
		&\frac{\sum_{m=1}^{m_1-1} l^r_n}{f^R_{\max}} \leq T_3, \label{e:busy_case_1_lagrangian_con_f3} 
		\end{align}
	\end{subequations}
\end{prob}
Furthermore, the dual problem of Problem \ref{p:busy_case_1_lower_transform} is given by
\begin{prob} \label{p:busy_case_1_dual}
	\begin{subequations}
		\begin{align}
		\max_{\lambda, \eta_1, \eta_2} & ~\mathcal{G} (\lambda, \eta_1, \eta_2) \notag \\
		\text{s.t.} \quad & ~\lambda \geq 0, \eta_1 \geq 0, \eta_2 \geq 0, 
		\end{align}
	\end{subequations}
\end{prob}

Since Problem \ref{p:busy_case_1_lower_transform} is convex and satisfies Slater's condition, strong duality holds between the primary problem and the dual problem, i.e. Problem \ref{p:busy_case_1_lower_transform} and Problem \ref{p:busy_case_1_dual}. Thus optimal solution of Problem \ref{p:busy_case_1_lower_transform} can be solved by equivalently solving Problem \ref{p:busy_case_1_dual}. Specifically, we solve Problem \ref{p:busy_case_1_lower_transform} by first evaluate the dual function $\mathcal{G} (\lambda, \eta_1, \eta_2)$ under a given set of $( \lambda, \eta_1, \eta_2 )$, and then obtain the optimal dual variables to maximize $\mathcal{G} (\lambda, \eta_1, \eta_2)$. 

To evaluate the dual function $\mathcal{G} (\lambda, \eta_1, \eta_2)$, for the optimal solution of Problem \ref{p:busy_case_1_lagrangian}, which is denoted as $(\tau_1^*,\tau_2^*,\tau_3^*,T_1^*,T_2^*,T_3^*)$, the following lemmas can be expected.

\begin{theo} \label{the:busy_optimal_tau1_tau2_T2}
	The optimal solution of $\tau_1$, $\tau_2$ and $T_2$ in Problem \ref{p:busy_case_1_lower_transform} can be expressed as
	\begin{align}
	\tau_1 &=\frac{d^s_{n_1}}{B \left(W_0 \left(\frac{\eta_2 h/\sigma^2-1}{e}\right)+1 \right)}, \label{e:busy_optimal_tau1} \\
	\tau_2 &=\frac{d^s_{n_2}}{B \left(W_0 \left(\frac{\eta_2 g/\sigma^2-1}{e}\right)+1 \right), \label{e:busy_optimal_tau2}} \\
	T_2 &=\max \left \{ \sqrt[3]{\frac{2 \kappa^r_0 \left( \sum_{n=n_1}^{n_2-1} l^s_n \right)^3}{\eta_2}}, \frac{\sum_{n=n_1}^{n_2-1} l^s_n}{f^R_{\max}} \right \}, \label{e:busy_optimal_T2}
	\end{align}
	in which $W_0(x)$ is the principal branch of the Lambert W function.
\end{theo}
\begin{IEEEproof}
	To calculate the optimal solution of $\tau_1$ and $\tau_2$, partial derivative of the function $\mathcal{L} (\tau_1, \tau_2, \tau_3, T_1, T_2, T_3, \lambda, \eta_1, \eta_2)$, which is written as $\mathcal{L}$ for simplification, is given as
	\begin{align}
	\frac{\partial \mathcal{L}}{\partial \tau_1} \bigg|_{\tau_1 =\tau_1^*} =
	\frac{\sigma^2}{h} \left(e^{\frac{d^s_{n_1}}{B \tau_1}}-1 \right) -\frac{\sigma^2 d^s_{n_1} e^{\frac{d^s_{n_1}}{B \tau_1}}}{Bh\tau_1} +\eta_2=0, \label{e:busy_case_1_KKT_dev_tau1} \\
	\frac{\partial \mathcal{L}}{\partial \tau_2} \bigg|_{\tau_2 =\tau_2^*} =
	\frac{\sigma^2}{g} \left(e^{\frac{d^s_{n_2}}{B \tau_2}}-1 \right) -\frac{\sigma^2 d^s_{n_2} e^{\frac{d^s_{n_2}}{B \tau_2}}}{Bg\tau_2} +\eta_2=0. \label{e:busy_case_1_KKT_dev_tau2}
	\end{align}
	Note that the expression of (\ref{e:busy_case_1_KKT_dev_tau1}) and (\ref{e:busy_case_1_KKT_dev_tau2}) is similar to (\ref{e:idle_KKT_dev_tau1}), thus the solution of $\tau_1$ and $\tau_2$ is similar to the proof of Theorem \ref{the:idle_optimal_tau} and is omitted for purpose of simplification.
	
	On the other hand, the partial derivative of $\mathcal{L}$ over $T_2$ is 
	\begin{align} \label{e:busy_case_1_KKT_dev_T2}
	\frac{\partial \mathcal{L}}{\partial T_2} \bigg|_{T_2 =T_2^*}
	=&-\frac{2 \kappa_0^r \left( \sum_{n=n_1}^{n_2-1} l^s_n \right)^3}{T_2^3} +\eta_2 \notag \\
	& \left \{ \begin{array}{ll}
	=0,  \frac{\sum_{n=n_1}^{n_2-1} l^s_n}{f^R_{\max}} < T_2^* \\
	>0,  \frac{\sum_{n=n_1}^{n_2-1} l^s_n}{f^R_{\max}} = T_2^*
	\end{array} \right \}. 
	\end{align}
	Utilizing (\ref{e:busy_case_1_KKT_dev_T2}), the optimal $T_2$ is derived.
\end{IEEEproof}

\begin{theo} \label{the:busy_optimal_tau3_T3}
	The optimal solution of $\tau_3$ and $T_3$ in Problem \ref{p:busy_case_1_lower_transform} can be expressed as
	\begin{align}
	\tau_3 &=\frac{d^r_{m_1}}{B \left(W_0 \left(\frac{(\lambda +\eta_1) g/\sigma^2-1}{e}\right)+1 \right), \label{e:busy_optimal_tau3}} \\
	T_3 &=\max \left \{ \sqrt[3]{\frac{2 \kappa^r_0 \left( \sum_{m=1}^{m_1-1} l^r_m \right)^3}{\lambda +\eta_1}}, \frac{\sum_{m=1}^{m_1-1} l^r_m}{f^R_{\max}} \right \}, \label{e:busy_optimal_T3}
	\end{align}
\end{theo}
\begin{IEEEproof}
	The partial derivative of $\mathcal{L}$ over $\tau_3$ is
	\begin{equation} \label{e:busy_case_1_KKT_dev_tau3}
	\frac{\partial \mathcal{L}}{\partial \tau_3} \bigg|_{\tau_3 =\tau_3^*} =
	\frac{\sigma^2}{g} \left(e^{\frac{d^r_{m_1}}{B \tau_3}}-1 \right) -\frac{\sigma^2 d^r_{m_1} e^{\frac{d^r_{m_1}}{B \tau_3}}}{Bg\tau_3} +\eta_1 +\lambda=0.
	\end{equation}
	Thereby (\ref{e:busy_optimal_tau3}) can be derived.
	The partial derivative of $\mathcal{L}$ over $T_3$ is
	\begin{align} \label{e:busy_case_1_KKT_dev_T3}
	\frac{\partial \mathcal{L}}{\partial T_3} \bigg|_{T_3 =T_3^*} =
	&-\frac{2 \kappa_0^r \left( \sum_{m=1}^{m_1-1} l^r_m \right)^3}{T_3^3} +\eta_1 +\lambda \notag \\
	& \left \{ \begin{array}{ll}
	=0,  \frac{\sum_{m=1}^{m_1-1} l^r_n}{f^R_{\max}} < T_3^* \\
	>0,  \frac{\sum_{m=1}^{m_1-1} l^r_n}{f^R_{\max}} = T_3^*
	\end{array} \right \}.
	\end{align}
	Utilizing (\ref{e:busy_case_1_KKT_dev_T3}), the optimal $T_3$ is derived.
\end{IEEEproof}

\begin{theo} \label{the:busy_optimal_T1}
	The optimal solution of $T_1$ in Problem \ref{p:busy_case_1_lower} can be expressed as
	\begin{equation} \label{e:busy_optimal_T1}
	T_1 =\max \left \{ \sqrt[3]{\frac{2 \kappa^s_0 \left( \sum_{n=1}^{n_1-1} l^s_n \right)^3}{\eta_2-\lambda}}, \frac{\sum_{n=1}^{n_1-1} l^s_n}{f^S_{\max}} \right \}, 
	\end{equation}
	and there exists $\eta_2 > \lambda$.
\end{theo}
\begin{IEEEproof}
	The partial derivative of $\mathcal{L}$ over $T_1$ is
	\begin{align} \label{e:busy_case_1_KKT_dev_T1}
	\frac{\partial \mathcal{L}}{\partial T_1} \bigg|_{T_1 =T_1^*} =
	&-\frac{2 \kappa_0^s \left( \sum_{n=1}^{n_1-1} l^s_n \right)^3}{T_1^3} +\eta_2 -\lambda \notag \\
	& \left \{ \begin{array}{ll}
	=0,  \frac{\sum_{n=1}^{n_1-1} l^s_n}{f^S_{\max}} < T_1^* \\
	>0,  \frac{\sum_{n=1}^{n_1-1} l^s_n}{f^S_{\max}} = T_1^*
	\end{array} \right \}.
	\end{align}
	Utilizing (\ref{e:busy_case_1_KKT_dev_T1}), the optimal $T_2$ is derived.
\end{IEEEproof}
To this end, the dual function $\mathcal{G} (\lambda, \eta_1, \eta_2)$ has been evaluated. In order to determine the optimal solution of the primary problem, i.e. Problem \ref{p:busy_case_1_lagrangian}, optimal $(\lambda, \eta_1, \eta_2)$ should be properly searched.
Since the dual function $\mathcal{G} (\lambda, \eta_1, \eta_2)$ is concave but may not be differentiable, one can converge to the optimal solution of these dual variables by resorting to the subgradient method \cite{subgradient_slides}.
In specific, the subgradient of $\mathcal{G} (\lambda, \eta_1, \eta_2)$ with respect to $(\lambda, \eta_1, \eta_2)$ is
\begin{align}
g_1(\lambda, \eta_1, \eta_2)=& T_0+T_3^*+ \tau_3^* - T_1^*, \\
g_2(\lambda, \eta_1, \eta_2)=& \frac{\sum_{n=n_2}^{N} l^s_n }{f^E_{\max}} +\frac{\sum_{m=m_1}^{M} l^r_m}{f^E_{\max}} +T_0 +T_3^* +\tau_3^* - T^R_{th}, \\
g_3(\lambda, \eta_1, \eta_2)=& \frac{\sum_{n=n_2}^{N} l^s_n }{f^E_{\max}} +T_1^* +\tau_1^* +T_2^* +\tau_2^* - T^S_{th}.
\end{align}
The dual variables $(\lambda, \eta_1, \eta_2)$ can be updated as
\begin{align}
\lambda(k+1) &=\left[ \lambda(k) -\alpha_k g_1(\lambda(k), \eta_1(k), \eta_2(k)) \right]^+, \\
\eta_1(k+1) &=\left[ \eta_1(k) -\alpha_k g_2(\lambda(k), \eta_1(k), \eta_2(k)) \right]^+, \\
\eta_2(k+1) &=\left[ \eta_2(k) -\alpha_k g_3(\lambda(k), \eta_1(k), \eta_2(k)) \right]^+,
\end{align}
where $\alpha_k$ is the positive step size at the $k$th iteration, and $(x)^+=\max \{x,0\}$.

Finally, to find optimal solution of $n_1$, $n_2$ and $m_1$ in Problem \ref{p:busy_case_1_upper}, are integer variables, traversal should be utilized to find the optimal solution.

\section{Conclusions}
\label{s:conclusion}
In this paper, we have investigated a MEC system in which one mobile device who has sequential tasks to complete is assisted by a cooperative node and a BS.
The cooperative node assist the mobile device on both task offloading and task computation.
Specifically, two cases are considered, which are
1) the cooperative node has no tasks to complete itself, and
2) the cooperative node has tasks to complete itself.
Our target is to minimize the total energy consumption of the mobile device and the cooperative node by adjusting the transmit duration in task offloading, CPU frequency in task computation and the index of the task to offload.
In the first case, a mix-integer non-convex problem is formulated.
To make the problem tractable, we decompose the problem into two level.
The lower level problem is convex and satisfies Slater's condition, therefore KKT conditions are utilized to reduce the problem into a one-dimension problem, for which bisection search is applied to find the optimal solution.
In the upper level problem, to find the optimal task index to offload with lower complexity, we exploit the monotonic structure of the task size to simplify the search.
In the second case, to guarantee the successful computation of the mobile device and the cooperative node, the schedule of uploading transmission is classified into three schemes.
Within each scheme, a mix-integer non-convex problem is formulated, which is decomposed into two levels.
In the lower level problem, semi-closed solution is found by KKT conditions, whereas in the upper level problem, traversal method is utilized to find the optimal offloading index.

\appendices

\section{Optimization Problems in Scheme 2 and Scheme 3 when the cooperative node has tasks to complete itself}
\label{s:appendix_problems}

In Scheme 2, Problem \ref{p:busy_first} can be transformed into the following form.
\begin{prob} \label{p:busy_case_2}
	\begin{subequations}
		\begin{align}
		\min_{\substack{n_1, n_2, m_1, \tau_1, \tau_2, \tau_3 \\ T_1, T_2, T_3, \tau_s}}
		&\frac{\sigma^2 \tau_1}{h} \left( e^{\frac{d^s_{n_1}}{\tau_1B}}-1 \right)+ \frac{\sigma^2 \tau_2}{g} \left( e^{\frac{d^s_{n_2}}{\tau_2B}}-1 \right) +\frac{\sigma^2 \tau_3}{g} \left( e^{\frac{d^s_{m_1}}{\tau_3B}}-1 \right) \notag \\
		&+\frac{\kappa_0^r \left( \sum_{m=1}^{m_1-1} l^r_m \right)^3}{T_3^2} +\frac{\kappa_0^s \left( \sum_{n=1}^{n_1-1} l^s_n \right)^3}{T_1^2} + \frac{\kappa_0^r \left( \sum_{n=n_1}^{n_2-1} l^s_n \right)^3}{T_2^2} \notag \\
		\text{s.t.} \qquad &T_1 + \tau_1 + T_2 + \tau_2 \leq T_0 +  T_2 + T_3, \label{e:busy_case_2_con_time} \\
		&\frac{\sum_{n=1}^{n_1-1} l^s_n}{T_1} \leq f^S_{\max}, \label{e:busy_case_2_con_f1} \\
		&\frac{\sum_{n=n_1}^{n_2-1} l^s_n}{T_2} \leq f^R_{\max}, \label{e:busy_case_2_con_f2} \\	
		&\frac{\sum_{m=1}^{m_1-1} l^r_n}{T_3} \leq f^R_{\max}, \label{e:busy_case_2_con_f3} \\
		&\sum_{n=n_2}^{N} l^s_n \leq f^E_{\max} \tau_s, \label{e:busy_case_2_con_mb} \\
		&\sum_{m=m_1}^{M} l^r_m \leq f^E_{\max} \left(T^R_{th} -T_c\right), \label{e:busy_case_2_con_rl} \\
		&\tau_s \leq T^S_{th} -T_1 -\tau_1 - T_2 -\tau_2. \label{e:busy_case_2_con_taus}
		\end{align}
	\end{subequations}
\end{prob}
In Problem \ref{p:busy_case_2},
\begin{equation}
T_c=\max\{T_1+\tau_1+T_2+\tau_2+\tau_s, T_0+T_2+T_3+\tau_3\}.
\end{equation}

In Scheme 3, Problem \ref{p:busy_first} can be transformed into the following form.
\begin{prob} \label{p:busy_case_3}
	\begin{subequations}
		\begin{align}
		\min_{\substack{n_1, n_2, m_1, \tau_1, \tau_2, \tau_3 \\ T_1, T_2, T_3, \tau_s, \tau_0}}
		&\frac{\sigma^2 \tau_1}{h} \left( e^{\frac{d^s_{n_1}}{\tau_1B}}-1 \right)+ \frac{\sigma^2 \tau_2}{g} \left( e^{\frac{d^s_{n_2}}{\tau_2B}}-1 \right) +\frac{\sigma^2 \tau_3}{g} \left( e^{\frac{d^s_{m_1}}{\tau_3B}}-1 \right) \notag \\
		&+\frac{\kappa_0^r \left( \sum_{m=1}^{m_1-1} l^r_m \right)^3}{T_3^2} +\frac{\kappa_0^s \left( \sum_{n=1}^{n_1-1} l^s_n \right)^3}{T_1^2} + \frac{\kappa_0^r \left( \sum_{n=n_1}^{n_2-1} l^s_n \right)^3}{T_2^2} \notag \\
		\text{s.t.} \qquad &T_1+ \tau_1 \leq T_0 + T_3, \label{e:busy_case_3_con_time1}\\
		&T_0 + T_3 + \tau_0 + \tau_3 \leq T_1 + \tau_1 + T_2, \label{e:busy_case_3_con_time2} \\
		&\frac{\sum_{n=1}^{n_1-1} l^s_n}{T_1} \leq f^S_{\max}, \label{e:busy_case_3_con_f1} \\
		&\frac{\sum_{n=n_1}^{n_2-1} l^s_n}{T_2} \leq f^R_{\max}, \label{e:busy_case_3_con_f2} \\	
		&\frac{\sum_{m=1}^{m_1-1} l^r_n}{T_3} \leq f^R_{\max}, \label{e:busy_case_3_con_f3} \\
		&\sum_{n=n_2}^{N} l^s_n \leq f^E_{\max} \tau_s, \label{e:busy_case_3_con_mb} \\
		&\sum_{m=m_1}^{M} l^r_m \leq f^E_{\max} \notag \\
		&\times \left(T^R_{th} -T_0 -T_3 -\tau_0 -\tau_3 -\tau_s \right), \label{e:busy_case_3_con_rl} \\
		&\tau_s \leq T^S_{th} -T_1 -\tau_1 - T_2 -\tau_2. \label{e:busy_case_3_con_taus}
		\end{align}
	\end{subequations}
\end{prob}
In this problem, $\tau_0$ is the time gap between the arrival of the mobile device's task on the cooperative node and the transmission of the cooperative node's task to the BS.
In the problem solution,  $\tau_0$ can be proved to be 0.

The optimization problem in Scheme 2 and Scheme 3 are in similar structure with Scheme 1, and the problem solution follows the same procedure.
Therefore, solution for optimization problems in Scheme 2 and Scheme 3 are left out.

\end{document}